\documentstyle[preprint,aps,prl,epsf]{revtex}

\begin{document}

\tightenlines

\renewcommand{\thefootnote}{\fnsymbol{footnote}}

%%%%%%%%%%%%%%%
%% Substitute your Pub number, month and year
%% in the following
%%%%%%%%%%%%%%%

\begin{flushright}
{\small
SLAC--PUB--8280\\
SCIPP 99/45\\
October 1999\\
(T/E)\\}
\end{flushright}

\vspace{.8cm}

%%%%% Title and Author Information:
%%
\begin{center}
{\large\bf
Search for Charmless Hadronic Decays of $B$ Mesons with the SLD
Detector \footnote{Work supported by
Department of Energy contract  DE--AC03--76SF00515 (SLAC).}}

\vspace{1cm}

The SLD Collaboration$^{\dagger}$\\
Stanford Linear Accelerator Center \\ 
Stanford University, Stanford, CA  94309\\

\end{center}

\vfill

%%%%% If your paper has an abstract, put it here:
%%
\begin{center}
{\large\bf
Abstract }
\end{center}

\begin{quote}
Based on a sample of approximately 500,000 hadronic $Z^0$
decays accumulated between 1993 and 1998, the SLD experiment
has set limits on 24 fully charged two-body and
quasi two-body exclusive charmless hadronic 
decays of $B^{+}$, $B^0$, and $B^0_s$ mesons.
The precise tracking capabilities of the SLD detector provided for the
efficient reduction of combinatoric backgrounds, yielding the most precise
available limits for ten of these modes.
\end{quote}

\vfill

%%%%%%%%%%%%%%%
\begin{center}
{\it Submitted to Physical Review D -- Rapid Communications} \\
\end{center}

\eject

%Physical Review Letters
%Title: Search for Charmless Hadronic Decays of B Mesons with the SLD Detector
%Authors: K.Abe, K.Abe, T.Akagi, N.J.Allen, W.W.Ash, et al.
%Corresponding author: Bruce Schumm
%Email: schumm@scipp.ucsc.edu
%Address: SCIPP, Natural Sciences II
%University of California
%Santa Cruz, CA 95064
%Phone number: 831-459-3034
%FAX number: 831-459-5777
%
%Article type: Letter
%
%Suggested section: L-1 Elementary particles and fields
%
%Suggested principal PACS No.:
%     13.25.Hw
%Additional PACS No(s).:  
%
%%%%%%%%%%%%%%%%%%%%%%%%%%%%%%
%
%Version 1.0      08/09/99
%
%%%%%%%%%%%%%%%%%%%%%%%%%%%%%%
%____________________________________________________________________________
%
%% For LaTeX 2.09 Phys Rev Letters

%\documentstyle[floats,aps,prl,epsf]{revtex}
%\begin{document}

% Bibliography Macros

% A useful Journal macro
\newcommand{\Journal}[4]{{#1} {\bf #2} (#4) #3}
\newcommand{\SLDNote}[2]{SLD Note #1, (#2)}
\newcommand{\SLDPhys}[2]{SLD Physics Note #1, (#2)}
\newcommand{\SLCPhys}[2]{SLC Physics Note #1, (#2)}
\newcommand{\SLACPub}[2]{SLAC--PUB--#1, (#2)}
\newcommand{\SLACRep}[2]{SLAC--Report--#1, (#2)}
\newcommand{\SLACtn} [3]{SLAC--TN--#1--#2, (#3)}
\newcommand{\Thesis} [2]{PhD Thesis, #1 (#2)}

% Some useful journal names
\newcommand{\ARNPS}{{\em Annual Review of Nucl. Part. Sci.}}
\newcommand{\CPC} {{\em Comput. Phys. Commun.}}
\newcommand{\NCA} {\em Nuovo Cimento}
\newcommand{\NIM} {\em Nucl. Instrum. Methods}
\newcommand{\NIMA}{{\em Nucl. Instrum. Methods} A}
\newcommand{\NPB} {{\em Nucl. Phys.} B}
\newcommand{\NP}  {{\em Nucl. Phys.}}
\newcommand{\PA}  {{\em Part. Accel.}}
\newcommand{\PL}  {{\em Phys. Lett.}}
\newcommand{\PLB} {{\em Phys. Lett.}  B}
\newcommand{\PR}  {{\em Phys. Rev.}}
\newcommand{\PRL} {\em Phys. Rev. Lett.}
\newcommand{\PRD} {{\em Phys. Rev.} D}
\newcommand{\ZPC} {{\em Z. Phys.} C}
\newcommand{\EPJ} {{\em The European Physical Journal} C}
\newcommand{\etal}{{\em et al.}}
\newcommand{\ibid}{{\em ibid.}}

%lets get all the modes down ones and for all

\newcommand{\ma}{$B^{0}\rightarrow \pi^{+}\pi^{-}$}
\newcommand{\mb}{$B^{0}\rightarrow K^{-}\pi^{+}$}
\newcommand{\mc}{$B^{0}\rightarrow K^{+}K^{-}$}
\newcommand{\md}{$B^0_{s}\rightarrow \pi^{+}\pi^{-}$}
\newcommand{\me}{$B^0_{s}\rightarrow K^{-}\pi^{+}$}
\newcommand{\mf}{$B^0_{s}\rightarrow K^{+}K^{-}$}
\newcommand{\mg}{$B^{+}\rightarrow\rho^{0}\pi^{+}$}
\newcommand{\mh}{$B^{+}\rightarrow\rho^{0}K^{+}$}
\newcommand{\mi}{$B^{+}\rightarrow K^{\ast 0}\pi^{+}$}
\newcommand{\mj}{$B^{+}\rightarrow K^{\ast 0}K^{+}$}
\newcommand{\mk}{$B^{+}\rightarrow\phi \pi^{+}$}
\newcommand{\ml}{$B^{+}\rightarrow\phi K^{+}$}
\newcommand{\mm}{$B^{0}\rightarrow \rho^{0}\rho^{0}$}
\newcommand{\mn}{$B^{0}\rightarrow \bar{K}^{\ast 0}\rho^{0}$}
\newcommand{\mo}{$B^{0}\rightarrow \bar{K}^{\ast 0}K^{\ast 0}$}
\newcommand{\mq}{$B^{0}\rightarrow \phi\rho^{0}$}
\newcommand{\mr}{$B^{0}\rightarrow \phi\bar{K}^{\ast 0}$}
\newcommand{\ms}{$B^{0}\rightarrow \phi\phi$}
\newcommand{\mt}{$B^0_{s}\rightarrow \rho^{0}\rho^{0}$}
\newcommand{\mv}{$B^0_{s}\rightarrow \bar{K}^{\ast 0}\rho^{0}$}
\newcommand{\mw}{$B^0_{s}\rightarrow \bar{K}^{\ast 0}K^{\ast 0}$}
\newcommand{\mx}{$B^0_{s}\rightarrow \phi\rho^{0}$}
\newcommand{\my}{$B^0_{s}\rightarrow \phi {K}^{\ast 0}$}
\newcommand{\mz}{$B^0_{s}\rightarrow \phi\phi$}

%\draft

%\preprint{SLAC-PUB-7798}
%\preprint{*Draft* Version 3.0}
%\preprint{Mar/18/99}

%\title{Search for Charmless Hadronic Decays of $B$ Mesons
%       with the SLD Detector}

%\setcounter{page}{2}

%\maketitle

%\begin{abstract}
%Based on a sample of approximately 500,000 hadronic $Z^0$
%decays accumulated between 1993 and 1998, the SLD experiment
%has set limits on 24 fully charged two-body and
%quasi two-body exclusive charmless hadronic
%decays of $B^{+}$, $B^0$, and $B^0_s$ mesons.
%The precise tracking capabilities of the SLD detector provided for the
%efficient reduction of combinatoric backgrounds, yielding the most precise
%available limits for ten of these modes.

%\end{abstract}

%\pacs{13.25.Hw}

%\narrowtext

The search for exclusive charmless decays of $B$ mesons is motivated
by the CKM-suppression of the $W$-boson mediated $b \rightarrow u$
transition, which suppresses the leading order weak decay
to charmless final states by a factor of 
$|V_{ub}|^2/|V_{cb}|^2 \simeq 10^{-2}$ relative to
that of charmed final states. 
%The total branching ratio of $B$
%mesons into charmless final states is expected to be a few percent
%in the Standard Model, with most of the individual modes contributing
%$\sim 10^{-4}$ or less~\cite{RareB}.
Thus, observation of
exclusive charmless modes with even modest branching fractions can
indicate the participation of heretofore unobserved
physical processes.

Recently, several results have increased the interest
in exclusive charmless $B$ meson decays.
The CLEO collaboration~\cite{CLEOobs}
%PRL 80: 3710
has improved their measurement of
the decay $B \rightarrow \eta K^{\ast}$,
with the measured branching fractions
 $Br(B^+ \rightarrow \eta K^{\ast +}) = (2.73^{+0.96}_{-0.82} \pm 0.50)
   \times 10^{-5} $
and
 $Br(B^0 \rightarrow \eta K^{\ast 0}) = (1.38^{+0.55}_{-0.44} \pm 0.17)
   \times 10^{-5}$
somewhat
above the expected range~\cite{THEORY}
of $(0.02 - 0.82) \times 10^{-5}$ and
$(0.01 - 0.89) \times 10^{-5}$, respectively.
In addition, the DELPHI collaboration has 
reported a measurement~\cite{DELPHIobs}
%W. Adam {\it et. el.}, Z. Phys. C {\bf 72}, 207 (1996).
of the combined mode
$Br(B^+ \rightarrow \rho^0 \pi^+ + K^{*0} \pi^+) = (17^{+12}_{-8} \pm 2)
\times 10^{-5}$, again somewhat higher 
than both the expected range~\cite{THEORY}
of $(0.4 - 2.0)  \times 10^{-5}$ and the corresponding
CLEO measurements~\cite{CLEOlim}
of $Br(B^+ \rightarrow \rho^0 \pi^+) = 
(1.5^{+0.5}_{-0.5} \pm 0.4) \times 10^{-5} $
and $Br(B^+ \rightarrow K^{*0} \pi^+) < 2.7 \times 10^{-5}. $

In this Letter, we present limits from the SLD detector
on several two, three,
and four-prong fully charged two-body and quasi-two-body final
states.
Although the $B$ meson sample
available to the SLD detector is fairly limited in comparison to
those produced at LEP, CESR, and the TEVATRON,
the excellent tracking and
{\it a priori} knowledge of the $B$ meson production point
admit limits competitive with those produced elsewhere.
Most limits presented here on four-prong final states, for which
combinatoric backgrounds are worst, are the first available.
In addition, the
cms energy available to experiments running at the
$Z^0$ pole allows the study of $B_s$ decays, which are
inaccessible to experiments running at the $\Upsilon(4S)$.

Search modes reported in this Letter include
$B^0, B^0_s \rightarrow P^+P^-$ (two-prong),
$B^+ \rightarrow P^+V^0$ (three-prong), and
$B^0, B^0_s \rightarrow V^0V^0$ (four-prong),
and their charge-conjugates,
where $P = \pi, K$ is a stable pseudoscalar meson and
$V = \rho^0, K^{\ast 0}, \phi$ is a vector meson resonance with
a sizeable branching fraction into two charged pseudoscalar
mesons (100\%, 66.7\%, and 
(49.1 $\pm$ 0.8)\%~\cite{PDG}, respectively).
%PDG Collaboration, Eur. Phys. J. {\bf C3}, 1 (1998).
The ability to fully reconstruct the decaying $B$
meson state, with precise momentum and vertex information
for each of the charged daughter tracks, provides an essential
constraint in the analysis; no attempt was made to search
for modes with one or more long-lived
final state neutral particles.

The SLD detector~\cite{SLDdet}
%R_b PRD article; see trackqual cuts reference
instruments the sole interaction region
%K. Abe {\it et. al.}, Phys. Rev. {\bf D53}, 1023 (1996).
of the SLAC Linear Collider (SLC). 
The luminous region of the SLC is an ellipsoid of dimensions
approximately 2 and 0.8 $\mu m$ in the horizontal ($x$)
and vertical ($y$)
directions perpendicular to the beam axis, and 700 $\mu m$
along the beam axis. Due to motion of the collision point,
however, the location of the luminous region is known to
only $\sim 7 \mu m$ in $x$ and $y$.

Charged particle tracks are reconstructed in the central
drift chamber (CDC) and the CCD-based pixel vertex detector (VXD)
in a uniform axial magnetic field of 0.6T.
Including the uncertainty in the location of the luminous
region (IP) of the SLC,
the VXD2 vertex detector,
in place through 1995,
exhibited an $r-\phi$ ($r-z$) impact parameter resolution
of $11 \mu m$ ($38 \mu m$) at high momentum, and
$71 \mu m$ ($80 \mu m$) at $p_{\perp} \sqrt{\sin\theta} =
1.0$ GeV/c~\cite{VXD2res}.
%K. Abe {\it et al.}, Nucl. Instr. and Meth. {\bf A409}, 243 (1998).
The corresponding resolution for the VXD3
vertex detector~\cite{VXD3res},
%K. Abe {\it et al.}, Nucl. Instr. and Meth. {\bf A409}, 243 (1998).
in place since 1996, is
$14 \mu m$ ($26.5 \mu m$) at high momentum, and
$33 \mu m$ ($33 \mu m$) at $p_{\perp} \sqrt{\sin\theta} =
1.0$ GeV/c. The combined CDC/VXD momentum resolution in the
plane perpendicular to the beam axis is
$\delta p_{\perp} / p_{\perp} = \sqrt{(.01)^2 +
(.0026 p_{\perp} /{\rm GeV}/{\rm c})^2}$. High momentum charged tracks
are reconstructed in the range $|\cos \theta| <  0.85$,
with an efficiency of 96\% for
$|\cos\theta| < 0.65$. %Different between VXDII and III?
%The energy and angle of electromagnetic and hadronic showers
%is measured in the Liquid Argon
%Calorimeter (LAC), and the Warm Iron Calorimeter (WIC).
A segmented Si-W forward calorimeter, with polar angle
acceptance between 23 mr and 68 mr,
is used to monitor the SLC luminosity via t-channel
Bhabha scattering.
%A Cerenkov Ring Imaging Detector (CRID), located just inside the LAC,
%provides dedicated particle
%identification information, but was not found to be helpful in
%this analysis.

The SLD accumulated an
integrated luminosity of 19.1 $pb^{-1}$ of $e^+e^-$
annihilation data at the $Z^0$ pole between 1993 and 1998.
Of this, 14.0 $pb^{-1}$ was taken with the upgraded VXD3
vertex detector in place.

The complete reconstruction of the fully-charged final state
provides a number of constraints which can be used to
discriminate between signal and potential background sources.
Candidate
track combinations must be consistent
with having arisen from a single vertex. This vertex is displaced
from the collision point by an average of $\sim$3mm, which is measured
with an average uncertainty of 
$60 \mu$m  for the search modes.
The point of closest approach of the
extrapolated vertex momentum resultant
to the SLD IP (`vertex impact parameter') 
must be consistent with zero.
The invariant mass of the tracks forming the vertex must
be consistent with that of
the B meson 
%(5.28 GeV/c$^2$ for $B^0$, $B^+$,
%or 5.37 GeV/c$^2$ for $B^0_s$),
and have a total momentum
consistent with known fragmentation properties. Tracks emerging
from the $B$-meson decay vertex should have a relatively small
opening angle, a large momentum,
and a relatively large impact parameter with
respect to the SLD IP. For quasi two-body modes involving
vector meson resonances
($B \rightarrow PV$ and $B \rightarrow VV$), two of the charged tracks must have
an invariant mass consistent with that of each resonance. In addition,
for $B \rightarrow PV$ modes, the decay angle $\theta_h$ of the $V$ state
with respect to its flight direction (`helicity angle') must
be consistent with the 
distribution $d\Gamma / d\Omega_h \propto \cos^2 \theta_h$ dictated by
%$\cos \theta_h$  distribution dictated by
angular momentum conservation.

%Candidate decays were reconstructed by considering all combinatoric
%combinations of two, three, and four tracks which pass track
%quality cuts [ZZZ]
%Standard qual cut reference
%and with total charge $0$ for $PP$ and $VV$
%candidates, and $\pm 1$ for $PV$ candidates.
%Refer to some paper of the SLD here
%In the initial phase of the reconstruction, tracks from candidate vertices
%were required to have
%mutual opening angles of less than $\pi/3$
%radians, with
%at least one of the tracks having a momentum of 4 GeV/c or
%greater,
%and to all be within the same hemisphere as defined by the thrust axis.
%The probability of the vertex fit to the candidate tracks
%was required to be greater than $1 \times 10^{-5}$, with a significance
%(separation from the SLD IP divided by the associated error) of
%greater than 0.6, and a total energy of at least 20\% of the beam
%energy. Assuming a pion (kaon) mass for all vertex tracks, the invariant
%mass was required to be less than 6.4 (greater than 4.7)
%GeV/c$^2$. For three- and four-track candidates, the pion-hypothesis
%mass was also required to be greater than 4.0 GeV/c$^2$. For three-track
%candidates, the lowest mass two-track combination, corresponding to
%the vector meson candidate,
%was required to
%have a pion-hypothesis mass of less than 1.5 GeV/c$^2$, and a
%kaon-hypothesis mass of greater than 0.4 GeV/c$^2$. After these
%D%cuts, a total of approximately 2000 two-track, 10,000 three-track,
%and 40,000 four-track candidates remained.

Candidate decays were reconstructed by considering all
combinations of two, three, and four tracks which pass track
quality cuts~\cite{SLDdet}
%K. Abe {\it et al.}, Phys. Rev. {\bf D53}, 1023 (1996).
and with total charge $0$ for $PP$ and $VV$
candidates, and $\pm 1$ for $PV$ candidates.
%In the initial phase of the reconstruction, tracks from candidate vertices
%were required to have
%mutual opening angles of less than $\pi/3$
%radians, with
%at least one of the tracks having a momentum of 4 GeV/c or
%greater,
%and to all be within the same hemisphere as defined by the thrust axis.
The invariant mass of the candidate decay was required to be above
5.05 GeV/c$^2$ (5.15 GeV/c$^2$) for $B^+$ and $B^0$ ($B^0_s$) modes.
The probability of the vertex fit to the candidate tracks
was required to be greater than $1.0\%$
($0.5\%$ for $B \rightarrow VV$ modes and $B \rightarrow PV$ with $V = \phi$),
with a significance
(separation from the SLD IP divided by the associated error) 
of greater than 1.0
($0.6$ for $B \rightarrow VV$ modes and $B \rightarrow PV$ with $V = \phi$).
The smallest impact parameter $D$, normalized to its corresponding
uncertainty, of
any track in the candidate vertex was required to be greater than 1.1
($0.6$ for $B \rightarrow VV$ modes and $B \rightarrow PV$ modes
with $V = \phi$).
The change in the vertex invariant mass
between the assumption of the nominal (kaon) mass and 
pion mass for all relevant tracks in the candidate vertex
was required to be less than 0.3 GeV/c$^2$ (1.2 GeV/c$^2$)
for the $B \rightarrow P K^{\pm}$ ($B \rightarrow K \rho^0$) modes;
this cut suppresses background vertices
which get an artificially large mass due to a mistaken mass hypothesis for
one or more tracks.
A second mass reconstruction quantity
${\mathcal{M}}_{VV}$, defined to be the sum of the absolute values of
the differences between the reconstructed and nominal masses of the
vector meson and $B$ meson candidates,
was required to be less than 0.6 GeV/c$^2$,
0.4 GeV/c$^2$, and 0.4 GeV/c$^2$ for the
$B \rightarrow \rho^0 \rho^0$,
$B \rightarrow \rho^0 V$ ($V \neq \rho^0$), and
\mw modes, respectively. For relevant modes, the reconstructed vector meson
masses were required to be in the ranges [$0.2-1.1$],
[$0.7-1.0$], and [$1.000-1.035$] GeV/c$^2$ for $V = \rho^0$,
$K^{\ast 0}$, and $\phi$. Finally, $|\cos \theta_h|$ was
required to be greater than 0.3 for $B \rightarrow PV$ modes.

To further suppress background, an {\it ad-hoc}
discriminator function was devised, and tuned to a sample
of Monte Carlo (MC) $Z^0 \rightarrow b {\overline b}$ events
approximately ten times that of data, and a sample of light
quark (udsc) events approximately four times that of data.
For $B \rightarrow PP$ modes, this function took the form
\begin{eqnarray*}
{\mathcal{F}}_{PP} =   a_{0}e^{-\frac{(m-M_{b})^{2}}{2(\delta
m)^{2}}} -a_{1}e^{\frac{-m}{m_{0}}}-a_{2}e^{-\frac{S}{3}}
-a_{3}e^{-\tilde{\lambda}}
+a_{4}e^{-\frac{\lambda}{0.3}}
-a_{5}e^{-\frac{P}{0.03}}-a_{6}e^{-\frac{D}{3}} \\
-a_{7}e^{-\frac{\tilde{I}}{\tilde{I}_{0}}}+a_{8}e^{-\frac{I}{I_{0}}}
-a_{9}e^{-\frac{X}{0.5}}+a_{10}e^{-\frac{\Delta M}{0.2\mbox{\tiny GeV}}},
\end{eqnarray*}
with $m$ the invariant mass of the candidate vertex,
$S$ the vertex significance, $\lambda$
the largest angle between any tracks belonging to the vertex
($\tilde{\lambda} = 1 / \lambda -0.9$), $P$ the vertex fit
probability, $D$ the minimum normalized impact parameter,
$I$ the vertex impact parameter
($\tilde{I}\times 1000+24=1/(I+0.001)$), and $X = E_{vert}/E_{beam}$
the scaled vertex energy. $\Delta M$ is the difference in
the vertex mass between the pion and kaon hypotheses, and
exploits the propensity for all tracks deriving from decays of
the various search modes to be at high momentum.
The parameters $a_i > 0$, $m_0$, $I_0$, and $\tilde{I}_0$ were tuned
separately for the individual search modes, while
$M_{b}$ was set to 5.28 GeV/c$^2$ for $B^{0}$ or $B^{+}$
and 5.37 GeV/c$^2$ for $B_{s}$.

For $B\rightarrow PV$ modes, the discriminator function took the form
\begin{eqnarray*}
{\mathcal{F}}_{PV} &= & \left . \mathcal{F}_{PP}\right |_{a_{10=0}} +
 a_{11}e^{-\frac{(m_{v}-M_{v})^{2}}{2(\delta
 m_{v})^{2}}}+(1-\cos (h\pi )),
\end{eqnarray*}
with $m_{v}$ the invariant mass of the vector meson candidate,
and $h = \cos \theta_h$.
The vector meson masses $M_{v}$ were set to 0.77, 0.89 and
1.02 GeV/c$^2$ for $\rho$, $K^{\ast 0}$ and $\phi$ candidates,
respectively, with corresponding widths $\delta m_v$ of
$0.1$,
$0.08$  and $0.006$ GeV/c$^2$, respectively.

For $B\rightarrow VV$ modes, the discriminator function took the form
\begin{eqnarray*}
{\mathcal{F}}_{VV} &= & \left . \mathcal{F}_{PP}\right |_{a_{10=0}} +
 a_{11}e^{-\frac{(m^{(1)}_{v}-M^{(1)}_{v})^{2}}{2(\delta
 m^{(1)}_{v})^{2}}}+ a_{12}e^{-\frac{(m^{(2)}_{v}-M^{(2)}_{v})^{2}}{2(\delta
 m^{(2)}_{v})^{2}}}+a_{13}e^{-\frac{{\mathcal{M}}_{VV}}{0.4\mbox{\tiny GeV}}},
\end{eqnarray*}
with vector meson candidates selected according to the track
partition yielding vector meson masses closest to those of the
search mode. 

The discriminator functions were tuned for the various search
modes by maximizing the separation between the SLD MC sample
(which contains no charmless hadronic $B$ decays) and
separately generated MC samples representing each individual
search mode. The signal region for each search mode was
then defined according to a cut on the output of the
corresponding discriminator function. 
%above which
%an event was considered to be a search mode candidate.
For each search mode,
the value of this cut was selected in an unbiased way
by minimizing the average expected MC
Poisson upper limit $\mathcal{P}$ according to
\begin{eqnarray*}
\mathcal{P} &=& \sum_{i=0}^{\infty}P(u,i)Br_{i}(\varepsilon )
\end{eqnarray*}
where $P(u,i)$ is the Poisson probability for 
finding $i$ background events given
an expectation of $u$, and $Br_{i}(\varepsilon )$
is the 90\% CL upper
limit for the branching ratio if $i$ events are found.
The expected signal efficiency
from the MC simulation at these
optimal points ranged between 24.8\% and 37.9\% for the
various search modes, with expected backgrounds
of between
0.0 and 0.48 events. The efficiency is that for all $B$ meson
signal events, regardless of whether the decay occurred in
the fiducial region of the detector, but does not take
into account the branching ratios into fully-charged two-body
final states for the vector mesons in the
$B \rightarrow PV$ and $B \rightarrow VV$ modes.

%The signal efficiencies, which are necessary for the determination
%of the branching ratio limits, must be determined from the SLD MC
%simulation, and thus are subject to modeling uncertainties.
%Given the high momentum of the tracks from search mode decays,
%as well as the need to reconstruct every track from the
%$B$ meson decay in order to estimate the $B$ meson mass,
%the signal efficiency simulation is particularly sensitive
%to the improper modeling of tracking efficiency and track
%parameter smearing. It should be noted that the kinematics of the
%two-body and quasi two-body signal
%mode decays are completely prescribed by momentum and energy
%conservation.

The signal efficiencies were determined from the SLD MC
simulation, and thus are subject to modeling uncertainties.
%particularly the improper
%modeling of tracking efficiency and track
%parameter smearing.
The efficiency of the SLD tracking system was constrained by
studying the track multiplicity distributions of inclusively
tagged $Z^0 \rightarrow b {\overline b}$ events, which are
identified with approximately 98\% purity by the SLD~\cite{ZVTOP}.
%David J. Jackson, Nucl. Instr. and Meth. {\bf A388}, 247 (1997).
The kinematic distributions of tracks from such events are
well constrained by measurements of $B$ meson decay at the
$\Upsilon (4S)$~\cite{U4Sdec}, as well as measurements of
heavy-quark associated multiplicity at the $Z^0$ pole~\cite{NB}.
The resulting comparison of the momentum dependence of the multiplicity
between inclusively tagged MC and data events
indicated a deficit of $\sim 5\%$
in the tracking efficiency below 1.5 (0.8) GeV/c for the
VXD2 (VXD3) data sample, leading to a
reduction in the estimated signal mode efficiency
of
$\delta \varepsilon / \varepsilon \simeq 1-2\%$.
%for the two-prong modes
%to $\delta \varepsilon / \varepsilon \simeq 2\%$ for the four-prong modes.
%For these events, the kinematic distributions of tracks originating
%from the decay of the heavy hadrons (roughly 1/2 of the observed
%multiplicity) are well constrained from measurements at the
%$\Upsilon (4S)$~\cite{U4Sdec}.
%M. S. Alam {\it et al.}, Phys. Rev. Lett. {\bf 49}, 357 (1982).
%The magnitude of the remaining multiplicity (associated with
%the heavy quark fragmentation process) is also well constrained
%by measurements at the $Z^0$ pole~\cite{NB}.
%N_b measurements
%K. Abe {\it et al.}, Phys. Lett. {\bf B386}, 475 (1996)
%P. Abreu {\it et al.}, Phys. Lett. {\bf B347}, 447 (1995)
%P. D. Acton {\it et al.}, Z. Phys. {\bf C53}, 539 (1992)
%The resulting comparison of the momentum dependence of the multiplicity
%between inclusively tagged MC and data events, which is thus
%relatively model-independent, indicated a deficit of $\sim 5\%$
%in the tracking efficiency below 1.5 (0.8) GeV/c for the
%VXD2 (VXD3) data sample. Due to the high average momentum of signal-mode
%decay tracks, the reduction in sample mode efficiency indicated by
%including this additional tracking inefficiency in the MC
%was small, varying from
%$\delta \varepsilon / \varepsilon \simeq 1\%$ for the 2-prong modes
%to $\delta \varepsilon / \varepsilon \simeq 2\%$ for the 4-prong modes.

The possibility of longitudinal polarization
of the vector mesons in the $VV$ decay modes has been considered.
A longitudinally
polarized vector meson will decay with a $\cos^2\theta_h$
distribution, with tracks
from vector mesons decaying with small $\theta$ tending to be
reconstructed less efficiently, due to the relatively low
momentum of the backward-going track, as well as the resulting
angular proximity of the two decay tracks. The signal MC assumes
50\% longitudinal polarization for the vector mesons from $VV$
decays. Assuming a uniform probability distribution between
0\% and 100\% polarization, the resulting relative systematic error
in the efficiency of the $VV$ decay mode reconstruction is
$\Delta \varepsilon / \varepsilon = 1.5\%$.

The momentum resolution at high momentum was studied by comparing the
width of the reconstructed mass peak between data and MC for a sample
of exclusively reconstructed $D^+ \rightarrow K^- \pi^+ \pi^+$ decays.
To account for the somewhat larger width observed in data, the MC
momentum distribution was smeared according to
$ {1 / p_{\perp}} \rightarrow  {1 /p_{\perp}} + {\rm Gaussian}, $
for a Gaussian width of 0.002 (0.001) (GeV/c)$^{-1}$
for the VXD2 (VXD3) data sample. The resulting change in the MC mass width,
for example for the $B^+ \rightarrow \rho^0 \pi^+$ search mode, is from
146 to 184 MeV/c$^2$ for the VXD2 data sample.
The resulting efficiency loss varied between
$\delta \varepsilon / \varepsilon \simeq 2-5\%$.
Smearing of the radial and longitudinal track origin parameters, constrained
by comparisons of $r-\phi$ and $r-z$ impact parameter distributions between
data and MC, yielded an additional efficiency loss of
$\delta \varepsilon / \varepsilon \simeq 2-4\%$.
As a cross check, after the inclusion of the above corrections
in the MC efficiency calculation,
the number of reconstructed
$D^+ \rightarrow K^- \pi^+ \pi^+$ decays is within $3\%$ of
the MC expectation, well within the experimental uncertainty
on the $D^+ \rightarrow K^- \pi^+ \pi^+$ branching fraction and
the $D^+$ production rate.

Application of the various search mode selection algorithms
to the full 1993-8 SLD data sample yielded a total of
four distinct candidate events ($E_1$ -- $E_4$) which populated
the signal regions of six separate search modes. The
events observed (background expected) 
in each of these modes were as follows:
event $E_1$ for $B^0 \rightarrow \rho^0 \rho^0$ (0.31);
events $E_1$, $E_2$ for $B^0 \rightarrow
{\overline K}^{\ast 0} \rho^0$ (0.49);
events $E_1$, $E_2$, $E_3$ for $B^0 \rightarrow
{\overline K}^{\ast 0} K^{\ast 0} (0.27)$;
event $E_4$ for $B^0 \rightarrow
\phi {\overline K}^{\ast 0} (0.14)$;
event $E_1$ for $B^0_s \rightarrow
{\overline K}^{\ast 0} \rho^0 (0.34)$; and
events $E_1$, $E_3$ for $B^0 \rightarrow
{\overline K}^{\ast 0} K^{\ast 0} (0.17)$.
For the remaining search modes, no events were seen.

Thirteen of fifteen MC events which passed the full selection criteria
for at least one of the search modes had at least one identified
track coming from a $B$ meson decay, with $B \rightarrow D \pi$
accounting for approximately one half of these. In four of the thirteen cases,
reconstructed rest mass missing due to undetected charged or neutral
particles was supplied by random fragmentation tracks.

Figure 1 shows the expected signal and background distributions as
a function of the discriminator output from the MC simulation of
a typical mode -- the $B^0 \rightarrow \rho^0 \rho^0$ mode for the
VXD-2 running period -- assuming a branching fraction
Br($B^0 \rightarrow \rho^0 \rho^0$) = $10^{-4}$. Figure 2 shows
the relative rate between data and MC of the inclusion of background
as the discriminator cut is relaxed, for the same sample. At a branching
ratio of $10^{-4}$, a clear signal is expected, while backgrounds
seem to be well modeled.

The mode for which the observed signal was
least likely to be accounted for by a statistical fluctuation
in the expected background was
$B^0 \rightarrow {\overline K}^{\ast 0} K^{\ast 0}$.
The Poisson
likelihood of an expected background of 0.27 events fluctuating
to three or more events is 0.27\%, but depends strongly
on the value of the expected background. 
A study for this mode similar to that of Fig. 2 yielded
%additional background was admitted into the signal region
%by loosening the discriminator function cut
%
an additional 11 events, compared
to a MC expectation of an additional 3 events.
%(the agreement was satisfactory for other modes).
%
%showed good
%agreement between the addition of data events into the
%sample region and the increase in the MC background
%expectation for most signal modes. However, for the
%$B^0 \rightarrow {\overline K}^{\ast 0} K^{\ast 0}$ mode,
%relaxing the discriminator function cut by an amount
%that increased the MC expectation to 3 events yielded
%an additional 11 events in the signal region, presumably
%all background.
Thus, for this mode there is reason to
believe that the background is underestimated, and so,
as for other modes, only an upper limit will be quoted.

The branching ratio upper limits $L$ are related to
the statistical upper limits $\alpha$ on the number of
observed events according to
$\alpha = S \cdot L;  \;\;
S = N_B \cdot \varepsilon,$
with $N_B$ the estimated number
of applicable $B$ meson decays, and $\varepsilon$
the estimated efficiency for reconstructing the given
signal mode. 
%Branching ratio limits are thus sensitive
%to systematic uncertainty in the sensitivity S.
The number of $B^+ (B^0)$ and $B_s^0$ meson decays
in the full SLD data sample is estimated from the
measured SLD sample luminosity and known $B$ meson
production rates
to be $(1.02 \pm 0.05) \times 10^{5}$ and
$(0.27 \pm 0.05) \times 10^{5}$, respectively.
It has been assumed that $(21.7 \pm 0.1)\%$ of hadronic
$Z^0$ decays involve primary $b$ quarks, and of these,
$(39.7^{+1.8}_{-2.2})\%$ are $B^0$ or $B^+$ decays,
and $(10.5^{+1.8}_{-2.2})\%$ are $B^0_s$ decays~\cite{PDG}.
%PDG again
The uncertainty in the signal mode efficiencies was
conservatively estimated to be the total difference in
the MC efficiency estimate with and without the extra momentum,
tracking efficiency, and track origin parameter
smearing. Including the additional $VV$ mode polarization systematic error,
as well as the MC statistical error of
$\Delta \varepsilon / \varepsilon \simeq 2 - 5 \%$, due
to the limited size of the generated signal mode samples,
the total modeling error was between 6 -- 10\% for all modes.

%Branching ratio upper limits, taking into account the
%sensitivities and their uncertainties, observed signal size,
%and expected background level, have been calculated in two
%ways for each signal mode. In the Bayesian approach~\cite{BAYS}
%% Per's ref 71
%the error-smeared Poisson probability is integrated
%to the value of the branching fraction which yields
%the chosen confidence limit; this value of the branching
%fraction is then chosen as the upper limit. In the
%Classical approach~\cite{CLASS}
%%Per's ref 72
%the two-sided interval of possible observations
%corresponding to the chosen confidence
%level is calculated as a function of the assumed branching
%fraction, and the corresponding limits are
%given by the range of branching fraction values for
%which the calculated interval includes the observed signal.

Table 1 exhibits the number of candidate events, expected background,
efficiency, sensitivity (S), and resulting 90\% CL
upper limits for both the Bayesian~\cite{BAYS}
and Classical~\cite{CLASS} approaches
for the 24 search modes. Each four-prong mode limit
presented here, with the exception of
$B^0 \rightarrow \phi {\overline K}^{\ast 0}$ and
$B^0 \rightarrow \phi \phi$, either improves upon the existing
limit~\cite{PDG}, or is the first available limit for the
given mode. Two of the two-prong $B^0_s$ modes 
($\pi^+ \pi^-$ and $K^- \pi^+$) are competitive with existing
limits~\cite{PDG}.
Furthermore, a comparison of the probabilty distribution
for the combination of the $\rho^0 \pi^+$ and $K^{\ast 0}\pi^+$ modes
with that implied by the DELPHI result
BR($B^+ \rightarrow \rho^0 \pi^+, K^{\ast 0} \pi^+) =
(1.7^{+1.2}_{-0.8} \pm 0.2) \times 10^{-4}$
yields only a 10\% probability that the two measurements are consistent
with the same central value.

In conclusion, the excellent tracking capabilities of the
SLD detector have enabled the SLD to establish a number
of unique or competitive limits on the decay of $B$ mesons
to exclusive charmless final states. In particular, most
of the four-prong quasi two-body limits presented here
are the most stringent limits available.
In addition, the SLD limits of
BR($B^+ \rightarrow \rho^0 \pi^+) < 0.83 \times 10^{-4}$ and
BR($B^+ \rightarrow K^{\ast 0} \pi^+) < 1.19 \times 10^{-4}$
(90\% CL) rule out a DELPHI observation of the sum of these
two modes~\cite{DELPHIobs}
%DELPHI paper again
in favor of more stringent limits from CLEO~\cite{CLEOlim}.
%CLEO paper again

We thank the SLAC accelerator department for
outstanding efforts on our behalf. This work was supported by the
U.S. Department of Energy and National Science Foundation, the UK Particle
Physics and Astronomy Research Council, the Istituto Nazionale di Fisica
Nucleare of Italy and the Japan-US Cooperative Research Project on High
Energy Physics.

%\narrowtext

\vfill \eject

\noindent
$^{\dagger}$The SLD Collaboration

%\author{\input sldauth}
%
% author list for inclusion in LaTeX documents
% using \author{} and \address{} commands
%
% Institution number definitions:
%
\begin{center}
\def\iADEL{$^{(1)}$}
\def\iAOMORI{$^{(2)}$}
\def\iBOLO{$^{(3)}$}
\def\iBRI{$^{(4)}$}
\def\iBRUN{$^{(5)}$}
\def\iBU{$^{(6)}$}
\def\iCINC{$^{(7)}$}
\def\iCOLO{$^{(8)}$}
\def\iCOLU{$^{(9)}$}
\def\iCSU{$^{(10)}$}
\def\iFERR{$^{(11)}$}
\def\iFRAS{$^{(12)}$}
\def\iILLI{$^{(13)}$}
\def\iJHU{$^{(14)}$}
\def\iLBL{$^{(15)}$}
\def\iLTU{$^{(16)}$}
\def\iMASS{$^{(17)}$}
\def\iMISSI{$^{(18)}$}
\def\iMIT{$^{(19)}$}
\def\iMOSCOW{$^{(20)}$}
\def\iNAGO{$^{(21)}$}
\def\iOREG{$^{(22)}$}
\def\iOXF{$^{(23)}$}
\def\iPADO{$^{(24)}$}
\def\iPERU{$^{(25)}$}
\def\iPISA{$^{(26)}$}
\def\iRAL{$^{(27)}$}
\def\iRUTG{$^{(28)}$}
\def\iSLAC{$^{(29)}$}
\def\iSOGA{$^{(30)}$}
\def\iSOONG{$^{(31)}$}
\def\iTENN{$^{(32)}$}
\def\iTOHO{$^{(33)}$}
\def\iUCSB{$^{(34)}$}
\def\iUCSC{$^{(35)}$}
\def\iUVIC{$^{(36)}$}
\def\iVAND{$^{(37)}$}
\def\iWASH{$^{(38)}$}
\def\iWISC{$^{(39)}$}
\def\iYALE{$^{(40)}$}

  \baselineskip=.75\baselineskip
\mbox{Kenji  Abe\unskip,\iNAGO}
\mbox{Koya Abe\unskip,\iTOHO}
\mbox{T. Abe\unskip,\iSLAC}
\mbox{I. Adam\unskip,\iSLAC}
\mbox{T.  Akagi\unskip,\iSLAC}
\mbox{H. Akimoto\unskip,\iSLAC}
\mbox{N.J. Allen\unskip,\iBRUN}
\mbox{W.W. Ash\unskip,\iSLAC}
\mbox{D. Aston\unskip,\iSLAC}
\mbox{K.G. Baird\unskip,\iMASS}
\mbox{C. Baltay\unskip,\iYALE}
\mbox{H.R. Band\unskip,\iWISC}
\mbox{M.B. Barakat\unskip,\iLTU}
\mbox{O. Bardon\unskip,\iMIT}
\mbox{T.L. Barklow\unskip,\iSLAC}
\mbox{G.L. Bashindzhagyan\unskip,\iMOSCOW}
\mbox{J.M. Bauer\unskip,\iMISSI}
\mbox{G. Bellodi\unskip,\iOXF}
\mbox{A.C. Benvenuti\unskip,\iBOLO}
\mbox{G.M. Bilei\unskip,\iPERU}
\mbox{D. Bisello\unskip,\iPADO}
\mbox{G. Blaylock\unskip,\iMASS}
\mbox{J.R. Bogart\unskip,\iSLAC}
\mbox{G.R. Bower\unskip,\iSLAC}
\mbox{J.E. Brau\unskip,\iOREG}
\mbox{M. Breidenbach\unskip,\iSLAC}
\mbox{W.M. Bugg\unskip,\iTENN}
\mbox{D. Burke\unskip,\iSLAC}
\mbox{T.H. Burnett\unskip,\iWASH}
\mbox{P.N. Burrows\unskip,\iOXF}
\mbox{R.M. Byrne\unskip,\iMIT}
\mbox{A. Calcaterra\unskip,\iFRAS}
\mbox{D. Calloway\unskip,\iSLAC}
\mbox{B. Camanzi\unskip,\iFERR}
\mbox{M. Carpinelli\unskip,\iPISA}
\mbox{R. Cassell\unskip,\iSLAC}
\mbox{R. Castaldi\unskip,\iPISA}
\mbox{A. Castro\unskip,\iPADO}
\mbox{M. Cavalli-Sforza\unskip,\iUCSC}
\mbox{A. Chou\unskip,\iSLAC}
\mbox{E. Church\unskip,\iWASH}
\mbox{H.O. Cohn\unskip,\iTENN}
\mbox{J.A. Coller\unskip,\iBU}
\mbox{M.R. Convery\unskip,\iSLAC}
\mbox{V. Cook\unskip,\iWASH}
\mbox{R.F. Cowan\unskip,\iMIT}
\mbox{D.G. Coyne\unskip,\iUCSC}
\mbox{G. Crawford\unskip,\iSLAC}
\mbox{C.J.S. Damerell\unskip,\iRAL}
\mbox{M.N. Danielson\unskip,\iCOLO}
\mbox{M. Daoudi\unskip,\iSLAC}
\mbox{N. de Groot\unskip,\iBRI}
\mbox{R. Dell'Orso\unskip,\iPERU}
\mbox{P.J. Dervan\unskip,\iBRUN}
\mbox{R. de Sangro\unskip,\iFRAS}
\mbox{M. Dima\unskip,\iCSU}
\mbox{D.N. Dong\unskip,\iMIT}
\mbox{M. Doser\unskip,\iSLAC}
\mbox{R. Dubois\unskip,\iSLAC}
\mbox{B.I. Eisenstein\unskip,\iILLI}
\mbox{I.Erofeeva\unskip,\iMOSCOW}
\mbox{V. Eschenburg\unskip,\iMISSI}
\mbox{E. Etzion\unskip,\iWISC}
\mbox{S. Fahey\unskip,\iCOLO}
\mbox{D. Falciai\unskip,\iFRAS}
\mbox{C. Fan\unskip,\iCOLO}
\mbox{J.P. Fernandez\unskip,\iUCSC}
\mbox{M.J. Fero\unskip,\iMIT}
\mbox{K. Flood\unskip,\iMASS}
\mbox{R. Frey\unskip,\iOREG}
\mbox{J. Gifford\unskip,\iUVIC}
\mbox{T. Gillman\unskip,\iRAL}
\mbox{G. Gladding\unskip,\iILLI}
\mbox{S. Gonzalez\unskip,\iMIT}
\mbox{E.R. Goodman\unskip,\iCOLO}
\mbox{E.L. Hart\unskip,\iTENN}
\mbox{J.L. Harton\unskip,\iCSU}
\mbox{K. Hasuko\unskip,\iTOHO}
\mbox{S.J. Hedges\unskip,\iBU}
\mbox{S.S. Hertzbach\unskip,\iMASS}
\mbox{M.D. Hildreth\unskip,\iSLAC}
\mbox{J. Huber\unskip,\iOREG}
\mbox{M.E. Huffer\unskip,\iSLAC}
\mbox{E.W. Hughes\unskip,\iSLAC}
\mbox{X. Huynh\unskip,\iSLAC}
\mbox{H. Hwang\unskip,\iOREG}
\mbox{M. Iwasaki\unskip,\iOREG}
\mbox{D.J. Jackson\unskip,\iRAL}
\mbox{P. Jacques\unskip,\iRUTG}
\mbox{J.A. Jaros\unskip,\iSLAC}
\mbox{Z.Y. Jiang\unskip,\iSLAC}
\mbox{A.S. Johnson\unskip,\iSLAC}
\mbox{J.R. Johnson\unskip,\iWISC}
\mbox{R.A. Johnson\unskip,\iCINC}
\mbox{T. Junk\unskip,\iSLAC}
\mbox{R. Kajikawa\unskip,\iNAGO}
\mbox{M. Kalelkar\unskip,\iRUTG}
\mbox{Y. Kamyshkov\unskip,\iTENN}
\mbox{H.J. Kang\unskip,\iRUTG}
\mbox{I. Karliner\unskip,\iILLI}
\mbox{H. Kawahara\unskip,\iSLAC}
\mbox{Y.D. Kim\unskip,\iSOGA}
\mbox{M.E. King\unskip,\iSLAC}
\mbox{R. King\unskip,\iSLAC}
\mbox{R.R. Kofler\unskip,\iMASS}
\mbox{N.M. Krishna\unskip,\iCOLO}
\mbox{R.S. Kroeger\unskip,\iMISSI}
\mbox{M. Langston\unskip,\iOREG}
\mbox{A. Lath\unskip,\iMIT}
\mbox{D.W.G. Leith\unskip,\iSLAC}
\mbox{V. Lia\unskip,\iMIT}
\mbox{C.Lin\unskip,\iMASS}
\mbox{M.X. Liu\unskip,\iYALE}
\mbox{X. Liu\unskip,\iUCSC}
\mbox{M. Loreti\unskip,\iPADO}
\mbox{A. Lu\unskip,\iUCSB}
\mbox{H.L. Lynch\unskip,\iSLAC}
\mbox{J. Ma\unskip,\iWASH}
\mbox{M. Mahjouri\unskip,\iMIT}
\mbox{G. Mancinelli\unskip,\iRUTG}
\mbox{S. Manly\unskip,\iYALE}
\mbox{G. Mantovani\unskip,\iPERU}
\mbox{T.W. Markiewicz\unskip,\iSLAC}
\mbox{T. Maruyama\unskip,\iSLAC}
\mbox{H. Masuda\unskip,\iSLAC}
\mbox{E. Mazzucato\unskip,\iFERR}
\mbox{A.K. McKemey\unskip,\iBRUN}
\mbox{B.T. Meadows\unskip,\iCINC}
\mbox{G. Menegatti\unskip,\iFERR}
\mbox{R. Messner\unskip,\iSLAC}
\mbox{P.M. Mockett\unskip,\iWASH}
\mbox{K.C. Moffeit\unskip,\iSLAC}
\mbox{T.B. Moore\unskip,\iYALE}
\mbox{M.Morii\unskip,\iSLAC}
\mbox{D. Muller\unskip,\iSLAC}
\mbox{V. Murzin\unskip,\iMOSCOW}
\mbox{T. Nagamine\unskip,\iTOHO}
\mbox{S. Narita\unskip,\iTOHO}
\mbox{U. Nauenberg\unskip,\iCOLO}
\mbox{H. Neal\unskip,\iSLAC}
\mbox{M. Nussbaum\unskip,\iCINC}
\mbox{N. Oishi\unskip,\iNAGO}
\mbox{D. Onoprienko\unskip,\iTENN}
\mbox{L.S. Osborne\unskip,\iMIT}
\mbox{R.S. Panvini\unskip,\iVAND}
\mbox{C.H. Park\unskip,\iSOONG}
\mbox{T.J. Pavel\unskip,\iSLAC}
\mbox{I. Peruzzi\unskip,\iFRAS}
\mbox{M. Piccolo\unskip,\iFRAS}
\mbox{L. Piemontese\unskip,\iFERR}
\mbox{K.T. Pitts\unskip,\iOREG}
\mbox{R.J. Plano\unskip,\iRUTG}
\mbox{R. Prepost\unskip,\iWISC}
\mbox{C.Y. Prescott\unskip,\iSLAC}
\mbox{G.D. Punkar\unskip,\iSLAC}
\mbox{J. Quigley\unskip,\iMIT}
\mbox{B.N. Ratcliff\unskip,\iSLAC}
\mbox{T.W. Reeves\unskip,\iVAND}
\mbox{J. Reidy\unskip,\iMISSI}
\mbox{P.L. Reinertsen\unskip,\iUCSC}
\mbox{P.E. Rensing\unskip,\iSLAC}
\mbox{L.S. Rochester\unskip,\iSLAC}
\mbox{P.C. Rowson\unskip,\iCOLU}
\mbox{J.J. Russell\unskip,\iSLAC}
\mbox{O.H. Saxton\unskip,\iSLAC}
\mbox{T. Schalk\unskip,\iUCSC}
\mbox{R.H. Schindler\unskip,\iSLAC}
\mbox{B.A. Schumm\unskip,\iUCSC}
\mbox{J. Schwiening\unskip,\iSLAC}
\mbox{S. Sen\unskip,\iYALE}
\mbox{V.V. Serbo\unskip,\iSLAC}
\mbox{M.H. Shaevitz\unskip,\iCOLU}
\mbox{J.T. Shank\unskip,\iBU}
\mbox{G. Shapiro\unskip,\iLBL}
\mbox{D.J. Sherden\unskip,\iSLAC}
\mbox{K.D. Shmakov\unskip,\iTENN}
\mbox{C. Simopoulos\unskip,\iSLAC}
\mbox{N.B. Sinev\unskip,\iOREG}
\mbox{S.R. Smith\unskip,\iSLAC}
\mbox{M.B. Smy\unskip,\iCSU}
\mbox{J.A. Snyder\unskip,\iYALE}
\mbox{H. Staengle\unskip,\iCSU}
\mbox{A. Stahl\unskip,\iSLAC}
\mbox{P. Stamer\unskip,\iRUTG}
\mbox{H. Steiner\unskip,\iLBL}
\mbox{R. Steiner\unskip,\iADEL}
\mbox{M.G. Strauss\unskip,\iMASS}
\mbox{D. Su\unskip,\iSLAC}
\mbox{F. Suekane\unskip,\iTOHO}
\mbox{A. Sugiyama\unskip,\iNAGO}
\mbox{S. Suzuki\unskip,\iNAGO}
\mbox{M. Swartz\unskip,\iJHU}
\mbox{A. Szumilo\unskip,\iWASH}
\mbox{T. Takahashi\unskip,\iSLAC}
\mbox{F.E. Taylor\unskip,\iMIT}
\mbox{J. Thom\unskip,\iSLAC}
\mbox{E. Torrence\unskip,\iMIT}
\mbox{N.K. Toumbas\unskip,\iSLAC}
\mbox{T. Usher\unskip,\iSLAC}
\mbox{C. Vannini\unskip,\iPISA}
\mbox{J. Va'vra\unskip,\iSLAC}
\mbox{E. Vella\unskip,\iSLAC}
\mbox{J.P. Venuti\unskip,\iVAND}
\mbox{R. Verdier\unskip,\iMIT}
\mbox{P.G. Verdini\unskip,\iPISA}
\mbox{D.L. Wagner\unskip,\iCOLO}
\mbox{S.R. Wagner\unskip,\iSLAC}
\mbox{A.P. Waite\unskip,\iSLAC}
\mbox{S. Walston\unskip,\iOREG}
\mbox{S.J. Watts\unskip,\iBRUN}
\mbox{A.W. Weidemann\unskip,\iTENN}
\mbox{E. R. Weiss\unskip,\iWASH}
\mbox{J.S. Whitaker\unskip,\iBU}
\mbox{S.L. White\unskip,\iTENN}
\mbox{F.J. Wickens\unskip,\iRAL}
\mbox{B. Williams\unskip,\iCOLO}
\mbox{D.C. Williams\unskip,\iMIT}
\mbox{S.H. Williams\unskip,\iSLAC}
\mbox{S. Willocq\unskip,\iMASS}
\mbox{R.J. Wilson\unskip,\iCSU}
\mbox{W.J. Wisniewski\unskip,\iSLAC}
\mbox{J. L. Wittlin\unskip,\iMASS}
\mbox{M. Woods\unskip,\iSLAC}
\mbox{G.B. Word\unskip,\iVAND}
\mbox{T.R. Wright\unskip,\iWISC}
\mbox{J. Wyss\unskip,\iPADO}
\mbox{R.K. Yamamoto\unskip,\iMIT}
\mbox{J.M. Yamartino\unskip,\iMIT}
\mbox{X. Yang\unskip,\iOREG}
\mbox{J. Yashima\unskip,\iTOHO}
\mbox{S.J. Yellin\unskip,\iUCSB}
\mbox{C.C. Young\unskip,\iSLAC}
\mbox{H. Yuta\unskip,\iAOMORI}
\mbox{G. Zapalac\unskip,\iWISC}
\mbox{R.W. Zdarko\unskip,\iSLAC}
\mbox{J. Zhou\unskip.\iOREG}

\it
  \vskip \baselineskip                   % \bigskip did not work
  \centerline{(The SLD Collaboration)}   % include collaboration name
  \vskip \baselineskip
  \baselineskip=.75\baselineskip   % shrink the interline spacing
\iADEL
  Adelphi University, Garden City, New York 11530, \break
\iAOMORI
  Aomori University, Aomori , 030 Japan, \break
\iBOLO
  INFN Sezione di Bologna, I-40126, Bologna, Italy, \break
\iBRI
  University of Bristol, Bristol, U.K., \break
\iBRUN
  Brunel University, Uxbridge, Middlesex, UB8 3PH United Kingdom, \break
\iBU
  Boston University, Boston, Massachusetts 02215, \break
\iCINC
  University of Cincinnati, Cincinnati, Ohio 45221, \break
\iCOLO
  University of Colorado, Boulder, Colorado 80309, \break
\iCOLU
  Columbia University, New York, New York 10533, \break
\iCSU
  Colorado State University, Ft. Collins, Colorado 80523, \break
\iFERR
  INFN Sezione di Ferrara and Universita di Ferrara, I-44100 Ferrara, Italy, \break
\iFRAS
  INFN Lab. Nazionali di Frascati, I-00044 Frascati, Italy, \break
\iILLI
  University of Illinois, Urbana, Illinois 61801, \break
\iJHU
  Johns Hopkins University,  Baltimore, Maryland 21218-2686, \break
\iLBL
  Lawrence Berkeley Laboratory, University of California, Berkeley, California 94720,
\break
\iLTU
  Louisiana Technical University, Ruston,Louisiana 71272, \break
\iMASS
  University of Massachusetts, Amherst, Massachusetts 01003, \break
\iMISSI
  University of Mississippi, University, Mississippi 38677, \break
\iMIT
  Massachusetts Institute of Technology, Cambridge, Massachusetts 02139, \break
\iMOSCOW
  Institute of Nuclear Physics, Moscow State University, 119899, Moscow Russia, \break
\iNAGO
  Nagoya University, Chikusa-ku, Nagoya, 464 Japan, \break
\iOREG
  University of Oregon, Eugene, Oregon 97403, \break
\iOXF
  Oxford University, Oxford, OX1 3RH, United Kingdom, \break
\iPADO
  INFN Sezione di Padova and Universita di Padova I-35100, Padova, Italy, \break
\iPERU
  INFN Sezione di Perugia and Universita di Perugia, I-06100 Perugia, Italy, \break
\iPISA
  INFN Sezione di Pisa and Universita di Pisa, I-56010 Pisa, Italy, \break
\iRAL
  Rutherford Appleton Laboratory, Chilton, Didcot, Oxon OX11 0QX United Kingdom, \break
\iRUTG
  Rutgers University, Piscataway, New Jersey 08855, \break
\iSLAC
  Stanford Linear Accelerator Center, Stanford University, Stanford, California 94309,
 \break
\iSOGA
  Sogang University, Seoul, Korea, \break
\iSOONG
  Soongsil University, Seoul, Korea 156-743, \break
\iTENN
  University of Tennessee, Knoxville, Tennessee 37996, \break
\iTOHO
  Tohoku University, Sendai 980, Japan, \break
\iUCSB
  University of California at Santa Barbara, Santa Barbara, California 93106, \break
\iUCSC
  University of California at Santa Cruz, Santa Cruz, California 95064, \break
\iUVIC
  University of Victoria, Victoria, British Columbia, Canada V8W 3P6, \break
\iVAND
  Vanderbilt University, Nashville,Tennessee 37235, \break
\iWASH
  University of Washington, Seattle, Washington 98105, \break
\iWISC
  University of Wisconsin, Madison,Wisconsin 53706, \break
\iYALE
  Yale University, New Haven, Connecticut 06511. \break

\rm
%
%  }   % end of address list
\end{center}
%\end{document}

\vfill
\eject

\begin{table}[tb]
 \centering
 \begin{tabular}{lcccccc}
%        &  &   &  &       &   (C)         &  (B)          \\
%        &  & S &  &       & UL (C)   &      UL (B)   \\
%       Mode    &$\varepsilon$ & $(\times 10^{-4})$ & Bckd& Data
%&      $(\times 10^{4})$&  $(\times 10^{4})$  \\ \hline
       Mode    &$\varepsilon$ & S $(\times 10^{-4})$ & Bckd& Data
&       UL (B; $\times 10^{4})$&  UL (C ; $\times 10^{4})$  \\ \hline
        \ma &.338 &$3.46 \pm .31$&0.03 & 0 & 0.69 & 0.67 \\
        \mb &.345 &$3.53 \pm .32$&0.14 & 0 & 0.67 & 0.66\\
        \mc &.341 &$3.49 \pm .28$&0.14 & 0 & 0.67 & 0.66\\  \hline
        \md &.379 &$1.02 \pm .16$&0.03 & 0 & 2.35 & 2.32\\
        \me &.335 &$0.91 \pm .15$&0.10 & 0 & 2.62 & 2.61\\
        \mf &.311 &$0.84 \pm .14$& 0.20 & 0 & 2.77 & 2.83\\ \hline
        \mg &.272 &$2.78 \pm .26$& 0.34 & 0 & 0.81 & 0.83\\
        \mh &.264 &$2.70 \pm .24$&0.41 & 0 & 0.83 & 0.86\\
        \mi &.285 &$1.94 \pm .17$&0.17 & 0 & 1.21 & 1.19\\
        \mj &.248 &$1.69 \pm .18$&0.17 & 0 & 1.39 & 1.38\\
        \mk &.301 &$1.51 \pm .12$&0.07 & 0 & 1.59 & 1.53\\
        \ml &.321 &$1.61 \pm .12$&0.14 & 0 & 1.47 & 1.44\\ \hline
        \mm &.263 &$2.76 \pm .24$& 0.31 & 1 & 1.57 & 1.36\\
        \mn &.253 &$1.76 \pm .15$& 0.49 & 2 & 3.30 & 2.86\\
        \mo &.304 &$1.42 \pm .12$& 0.27 & 3 & 5.27 & 4.69\\
        \mq &.298 &$1.53 \pm .13$& 0.14 & 0 & 1.58 & 1.56\\
        \mr &.295 &$1.01 \pm .08$& 0.14 & 1 & 4.34 & 3.84\\
        \ms &.393 &$0.74 \pm .05$& 0.00 & 0 & 3.37 & 3.21\\ \hline
        \mt &.277 &$0.77 \pm .13$& 0.27 & 0 & 3.06 & 3.20\\
        \mv &.272 &$0.50 \pm .09$& 0.34 & 1 & 8.52 & 7.67\\
        \mw &.265 &$0.33 \pm .05$& 0.17 & 2 & 18.21 & 16.81\\
        \mx &.290 &$0.39 \pm .06$& 0.07 & 0 & 6.24 & 6.17\\
        \my &.265 &$0.24 \pm .04$& 0.14 & 0 & 10.02 & 10.13\\
        \mz &.308 &$0.21 \pm .03$& 0.00 & 0 & 12.11 & 11.83\\
 \end{tabular}
 \caption[TABLE I]
        {Summary of efficiency ($\varepsilon$), sensitivity
(S), expected background, number of events in the signal region,
Classical (C) and Bayesian (B) 90\% CL Upper Limit for the 24 search modes.
Note that the sensitivities (but
not the efficiencies) take account of the branching fraction
for $\phi$ or $K^{\ast 0}$
into a fully-charged two
body final state, where applicable.
}
        \label{tab:summary}
\end{table}  

\vfill
\eject

\begin{figure}[t]
%\figurebox{20pc}{15pc}{} % to have a box alone
\epsfxsize=23pc % will enlarge or reduce the postscript figures based on the
\epsfbox{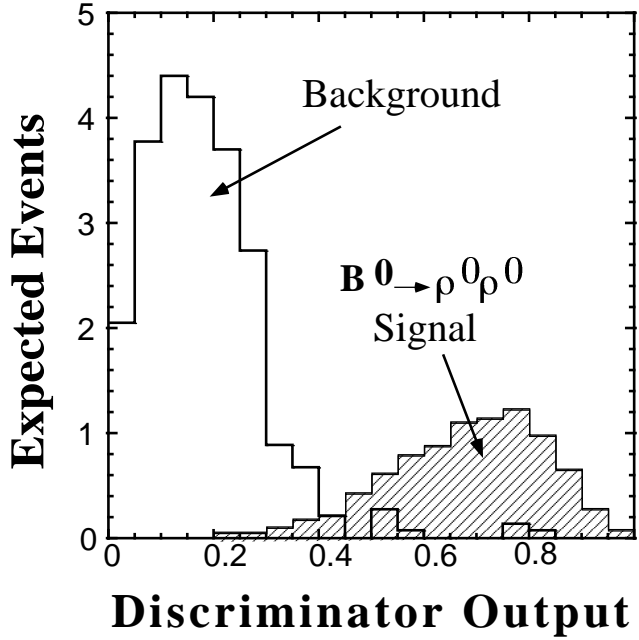} % postscript image file name
\caption{Expected signal (shaded) and background for the
$B^0 \rightarrow \rho^0 \rho^0$ channel, as a function of the discriminator
output. The plot shown is from the VXD-3 period Monte Carlo, assuming
a branching fraction 
Br($B^0 \rightarrow \rho^0 \rho^0$) = $10^{-4}$. \label{fig:bdback}}
\end{figure}

\begin{figure}[t]
%\figurebox{20pc}{15pc}{} % to have a box alone
\epsfxsize=21pc % will enlarge or reduce the postscript figures based on the
%\epsfbox{bdec_sig.ps} % postscript image file name
\epsfbox{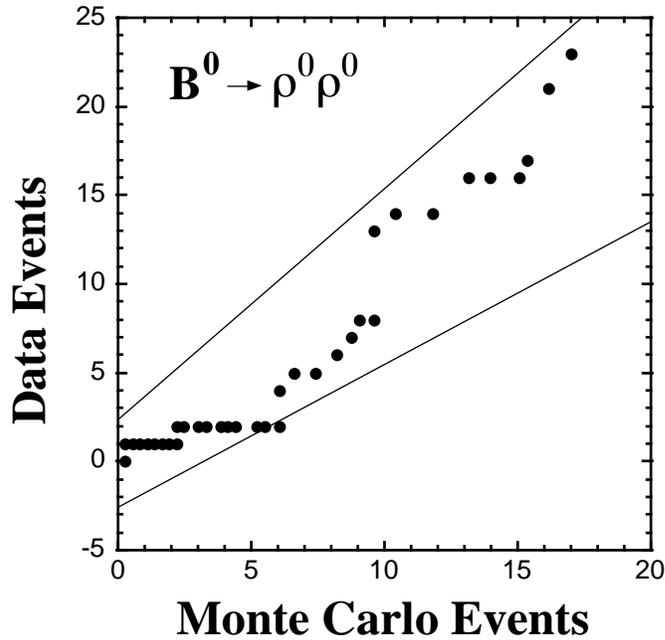} % postscript image file name
\caption{Comparison between VXD-3 data and Monte Carlo
of the rate of introduction of background into the
$B^0 \rightarrow \rho^0 \rho^0$ sample as the discriminator function cut
is relaxed. The lines represent the upper and lower 90\% CL limits under
the assumption that the Monte Carlo accurately 
models the background. \label{fig:bdsig}}
\end{figure}

\end{document}